\documentclass[11pt]{article}    
\usepackage[letterpaper,margin=1in]{geometry}

\usepackage{cite} 
\usepackage{url}  
\usepackage{ifthen}  
\usepackage{multicol}   
\usepackage[utf8]{inputenc} 
\usepackage{authblk}
\usepackage{graphicx}
\usepackage{amsmath,amsfonts,amssymb}
\usepackage{epsfig}
\usepackage{enumerate}
\usepackage{multirow}
\usepackage{algorithm,algorithmic}
\usepackage{bm}
\usepackage{hhline}
\usepackage{array}
\newcolumntype{k}[1]{%
>{\raggedleft\hspace{0pt}}p{#1}}%
\newcolumntype{x}[1]{%
>{\centering\hspace{0pt}}p{#1}}%

\newcommand{\eq}[1]{(\ref{#1})}
\graphicspath{{figs//}}

\newcommand{\hide}[1]{}

\newcommand{\bz}{{\textbf{z}}}

\begin{document}

\title{Comparing Nonparametric Bayesian Tree Priors for Clonal Reconstruction of Tumors}
 
\author[1]{Amit G. Deshwar}
\author[2]{Shankar Vembu}
\author[1,2,3,4,5]{Quaid Morris}
\affil[1]{Edward S. Rogers Sr. Department of Electrical and Computer Engineering, University of Toronto}
\affil[2]{Donnelly Center for Cellular and Biomolecular Research, University of Toronto}
\affil[3]{Department of Molecular Genetics, University of Toronto}
\affil[4]{Banting and Best Department of Medical Research, University of Toronto}
\affil[5]{Department of Computer Science, University of Toronto}
\date{}

\maketitle

\begin{abstract}      
Statistical machine learning methods, especially nonparametric
Bayesian methods, have become increasingly popular to infer clonal
population structure of tumors. 
Here we describe the treeCRP, an extension of the Chinese restaurant process (CRP),
a popular construction used in nonparametric mixture models, to infer
the phylogeny and genotype of major subclonal lineages represented in
the population of cancer cells. We also propose new split-merge
updates tailored to the subclonal reconstruction problem
that improve the mixing time of Markov chains. 
In comparisons with the tree-structured stick breaking prior used in
PhyloSub, we demonstrate superior mixing and running time using the
treeCRP with our new split-merge procedures.
We also show that given the same number of samples, TSSB and treeCRP have similar ability to recover the subclonal structure of a
tumor.

\end{abstract}

\section{Introduction}
The clonal theory of cancer posits that tumors contain 
multiple, genetically diverse subclonal populations of cells
that evolved from a single progenitor population through
successive waves of expansion and selection \cite{nowell1976}.  
Recent genetic analyses of tumor subpopulations support
this theory\cite{Gerlinger12,hughes2014clonal}. These analyses also identify characteristic
 driver mutations involved in cancer development and progression \cite{Hanahan11} and provide insight into understanding and predicting treatment response \cite{aparicio2013implications}.  Understanding this intratumor genotype heterogeneity is especially important because different subclonal populations have different abilities to metastasize and resist treatment \cite{ding2012clonal,Gerlinger12}.
These somatic mutations are detected through high-throughput sequencing of tumor and normal tissue; and can be broadly divided into two types: Simple Somatic Mutations (SSMs) consisting of substitutions and small insertions / deletions, and Copy Number Variations (CNVs) resulting from larger structural changes.

Current, widely-used high-throughput sequencing (HTS) technology generates short reads that rarely span multiple SSM loci, so in almost all cases only the \emph{variant allele frequency (VAF)}, \emph{i.e.}, the proportion of reads containing the variant, are available for individual SSMs.
These VAFs have been used to partially reconstruct the tumor subpopulations \cite{Mullighan08,Navin10,Marusyk10,nik12,Gerlinger12,Schuh12,Carter13,Landau13,trap,phylosub,pyclone,clomial,cloneHD,recBTP}.
However, surprisingly, these VAFs can be used to completely subpopulation genotypes in some cases, by reconstructing the evolutionary history of the subpopulations \cite{phylosub, trap, recBTP}; SSM VAFs from multiple tumor samples improve this reconstruction\cite{clomial, phylosub,cloneHD}.

These evolution-based subclonal reconstruction methods use a specific tree representation in which mutations are assigned to both internal and leaf nodes.
This representation excludes tree inference methods, like hierarchical clustering or the nested Chinese restaurant process \cite{blei2010nested} that assign observations (mutations) only to leaf nodes. To our knowledge only two tree-based statistical models have been described that i) allow mutations to be assigned to internal nodes and ii) are \emph{non-parametric}, i.e., that do not require pre-specification of the number of nodes. PhyloSub\cite{phylosub} has previously applied the tree-structured stick-breaking (TSSB) prior\cite{AdamsGJ10} to this problem. Here, we derive a version of the tree-Chinese Restaurant Process (treeCRP)\cite{MeedsRZR08} for subclonal reconstruction and new associated split-merge MCMC updates. We compare the two models in terms of their sampling efficiency and accuracy in subclonal reconstruction.

In the next section we provide an overview of the subclonal reconstruction problem.
The remainder of this paper consists of a formal description of the treeCRP model and the results from a series of empirical comparisons of the TSSB model against several treeCRP variants.

\section{Methods}
\subsection{Subclonal Reconstruction}
\begin{figure}[!ht]
\centering
\includegraphics[height=0.9\textheight]{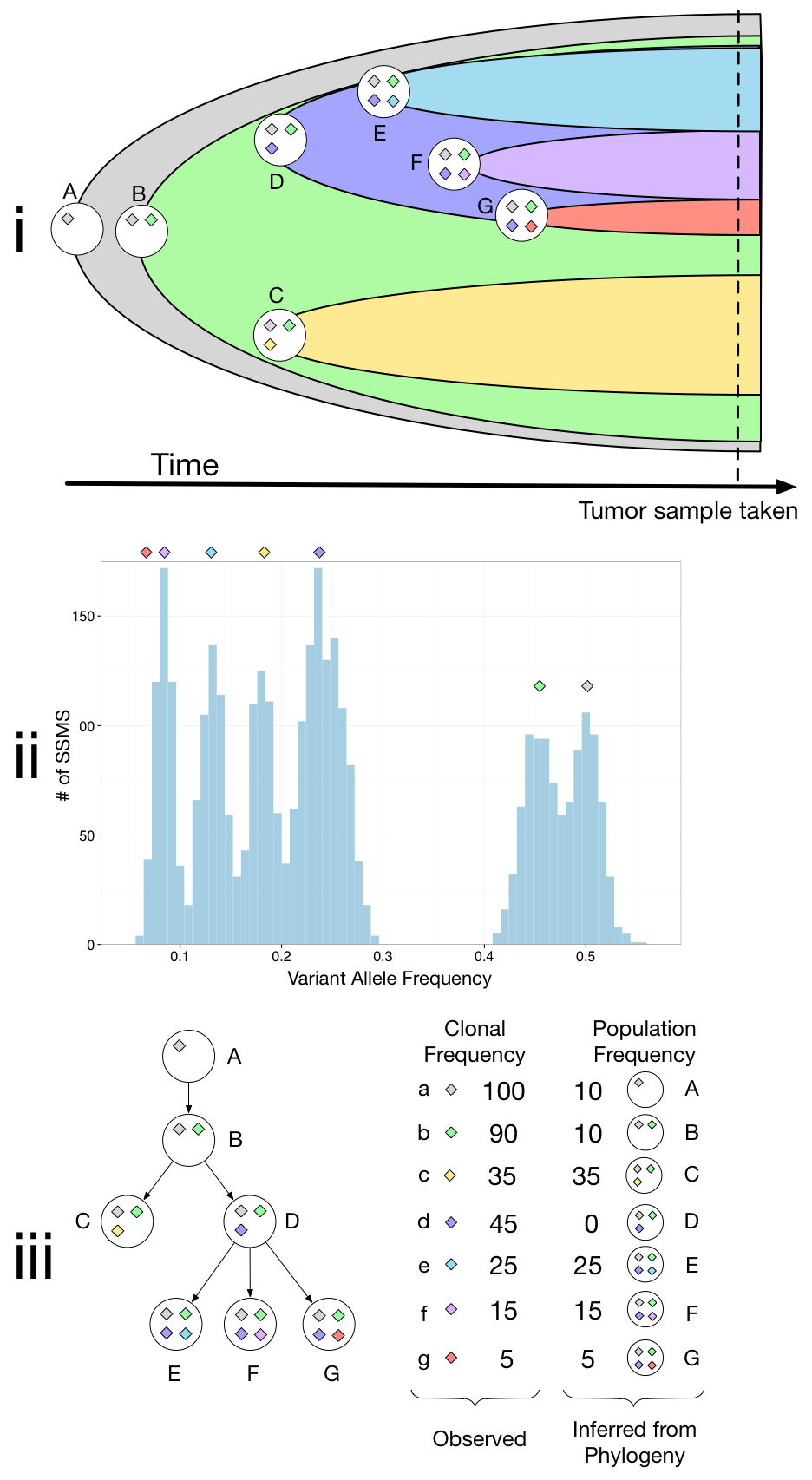}
\caption{The development of intratumor heterogeneity (i), the resulting distribution of variant allele frequencies (ii) and the desired output of subclonal inference (iii)}
\label{fig:clonal}
\end{figure}

Figure \ref{fig:clonal} provides an illustrative overview of the assumed process of tumor evolution and the task of subclonal reconstruction. Panel (i) of this figure shows a visualization of the evolution of a tumor over time as noncancerous cells (grey) are replaced by, at first, one clonal cancerous population (green) which then further develops into multiple cancerous subpopulations. Tumor cells define new subpopulations by acquiring new oncogenic mutations that allow their descendants to expanding relative to the other tumor subpopulations. Each circle in Panel (i) refers to a subpopulation.  We associate each subpopulation with the set of shared somatic mutations (shown as a diamond) that distinguish it from its parent subpopulation. However, each subpopulation also inherits all of its parent's mutations; as such, mutations may be present in multiple subpopulations. We define the \emph{subclonal lineage} of a mutation as the set of all subpopulations that contain it. For example, the subclonal lineage corresponding to the blue diamond includes the subpopulation (D) associated with that set of mutations and all decedent subpopulations (E,F,G). 
For clarity, and to highlight the link between subpopulations and their set of subpopulation-defining mutations, we will use the corresponding lower-case letter to refer to these mutations. For example, we will use \emph{d} to refer to the set of mutations represented by the blue diamond. 

Mutation sets, and their associated subpopulations, are defined by analyzing the population frequencies of somatic mutations detected in a tumor sample.
In the simple case that we consider here, SSMs occur in one copy of diploid regions of the genome; allowing one to estimate the \emph{clonal frequency} (i.e. the proportion of the sampled cells with the mutation) of a mutation by simply doubling its variant allele frequency, i.e., the proportion of reads mapping to the mutated locus that contain the SSM.  See Deshwar \emph{et al.} \cite{phylowgsarxiv} for the case when SSMs occur in non-diploid sections of the genome. 
Panel (ii) shows an example histogram of the SSM VAFs found in a heterogeneous tumor sample. Each subpopulation is defined by both the small number of oncogenic `driver' mutations that cause rapid expansion but also a larger number of `passenger' mutations acquired before the driver mutation(s) through errors in DNA replication (even noncancerous cells accumulate somatic mutations at a rate of $1.1$ per cell division \cite{behjati2014genome}). 
When a subpopulation expands, both the driver and the passenger SSMs increase in clonal frequency, and so have essentially identical frequencies. Due to sampling noise in the measurement of the VAFs, these mutation sets correspond to clusters (or modes) in the VAF distribution.  The central VAF of a particular cluster is determined by the population frequency of its subclonal lineage.
It is important to note that a given VAF cluster need not correspond to a subpopulation that is currently present in the tumor. For example, in Panel (ii), there is a VAF mode corresponding to mutation set \emph{d} even though subpopulation D has a population frequency of $0\%$ in the tumor sample. 
Only methods that attempt to reconstruct phylogenies (shown as panel (iii)), such as PhyloSub \cite{phylosub} and rec-BTP \cite{recBTP}, can detect when `vestigial' VAF clusters correspond to historical subpopulations that are no longer present in the sample.

\subsection{Our Approach}
We use a directed tree to represent the evolutionary relationship among the tumor subpopulations. Each node in the tree represents a subpopulation (either currently in the sample or that existed at some point in the tumor development) and the links connect parental subpopulations to their direct descendants. The set of SSMs assigned to a node are the defining set for the node's associated subpopulation. The subclonal lineage of an SSM consists of the subpopulation it is assigned to and that population's descendants. Each node $i$ is also assigned a frequency $\phi_i \in [0,1]$ which is the inferred clonal frequency of the SSMs in the node. The population frequency of the node's subpopulation, $\eta_i$, is the difference between the node's clonal frequency and the sum of the clonal frequencies of the node's children, \emph{i.e.}, $\eta_i = \phi_i - \sum_{j \in \mathcal{C}(i)} \phi_j$ where $\mathcal{C}(i)$ is the set of the indices of the children of node $i$.
The complete set of SSMs present in a subpopulation is the union of the SSMs assigned to it and those of all its ancestral nodes.

\subsection{Dirichlet process mixture models}
The treeCRP is derived from the Dirichlet process mixture model (DPMM) which we introduce here.
Consider the problem of clustering $N$ data objects $\{x_i\}_{i=1}^N$ using a Bayesian finite mixture model of $K$ components (clusters) with the following generative process \cite{Teh2010}:
\begin{equation}\label{eqn:fmm}
\begin{aligned}
\bm{\omega}  \sim \text{Dirichlet}(\alpha/K,\ldots,\alpha/K)\ ; \quad
& z_{i}  \sim \text{Multinomial}(\bm{\omega})\ ; \quad
\phi_{k} \sim  H\ ; \quad
& x_{i}  \sim  F(\phi_{z_i}) \ ,
\end{aligned}
\end{equation}
where $\bm{\omega}$ are the non-negative mixing weights such that $\sum_{k=1}^K \omega_k=1$, $\alpha$ is the concentration parameter of the symmetric Dirichlet prior placed on the mixing weights, $z_i \in \{1,\ldots,K\}$ is the cluster assignment variable, $H$ is the prior distribution from which the component parameters $\{\phi_k\}$ are drawn, $F(\phi)$ is the component distribution parameterized by $\phi$. The finite mixture model can be extended to a model with an infinite number of mixture components by replacing the Dirichlet prior with a Dirichlet process (DP) prior resulting in what is known as the DPMM \cite{Antoniak74}.
Unlike finite mixture models, DPMMs automatically estimate the number of components from the data thereby circumventing the problem of fixing the number of components \emph{a priori}. The Chinese Restaurant Process (CRP) provides a method to draw samples from a Dirichlet process. In this construction, an observation $x_i$ is assigned to an existing cluster $k$ with probability proportional to the number of objects $N^i_k$ in that cluster, excluding $x_i$. A new cluster $K+1$ is created with probability proportional to the concentration parameter. More formally,
\begin{equation}
\label{eqn:crp}
\begin{aligned}
p(z_i=k \mid \bz_{\backslash i}, \alpha ) & = \frac{N_k^{-i}}{N+\alpha-1} \ , \forall k \in \{1,\ldots,K\} \ ; \\
p(z_i=K+1 \mid \bz_{\backslash i}, \alpha) & = \frac{\alpha}{N+\alpha-1} \ ,
\end{aligned}
\end{equation}
where $\bz_{\backslash i} = \{z_1,\ldots,z_{i-1},z_{i+1},\ldots,z_N\}$.
The generative process for infinite mixture models using the Chinese restaurant process is:
\begin{equation}
\label{eqn:dpmm}
\begin{aligned}
z_{i}   \sim \text{CRP}(\alpha;\bz_{\backslash i}) \ ; \quad
\phi_{k} \sim  H \ ; \quad
x_{i}   \sim  F(\phi_{z_i}) \ .
\end{aligned}
\end{equation}

\hide{
An alternative view of the above generative process  produces component parameters $\{\tilde{\phi}_i\}$ by drawing samples from $\mathcal{G}$ resulting in the following generative process:
\begin{equation}\label{eqn:dpmm1}
\begin{aligned}
\mathcal{G}  \sim \mathrm{DP}(\alpha,H) \ ; \quad
\tilde{\phi}_{i}  \sim \mathcal{G} \ ; \quad
x_{i}  \sim  F(\tilde{\phi}_{i}) \ .
\end{aligned}
\end{equation}
Note that in the above process every object $\{x_i\}_{i=1}^N$ is associated with a component parameter $\{\tilde{\phi}_i\}_{i=1}^N$ and that all objects assigned to the same cluster will have the same component parameter. In other words, multiple elements in the set $\{\tilde{\phi}_i\}_{i=1}^N$ will take on the same value from the set $\{\phi_k\}_{k=1}^K$ of unique parameters. 
}

\subsection{Tree-structured Chinese restaurant process}
The Chinese restaurant process construction \eq{eqn:crp} described above can be used to produce a \emph{flat} clustering of objects, where the clusters are independent of each other. Meeds \emph{et al.} \cite{MeedsRZR08} extended this construction for \emph{relational} clustering that produces a clustering of objects where the clusters are connected to form a rooted tree.

In the tree-structured Chinese restaurant process (treeCRP), initially, a tree consists of a single root node (cluster) with all the data objects assigned to it. Subsequently,  an object $x_i$ is assigned to an existing node $k$ with probability proportional to the number of objects $N^i_k$ in that node, excluding $x_i$. A new node is $K+1$ is created as a child node to one of the $K$ existing nodes in the tree with probability proportional to $\alpha/K$. More formally,
\begin{equation}
\label{eqn:tcrp}
\begin{aligned}
p(z_i=k \mid \bz_{\backslash i}, \alpha ) & = \frac{N_k^{-i}}{N+\alpha-1} \ , \forall k \in \{1,\ldots,K\}  \ ; \\
p(z_i=K+1 \mid \bz_{\backslash i}, \alpha) & = \frac{1}{K}\left(\frac{\alpha} {N+\alpha-1}\right) \ .
\end{aligned}
\end{equation}

\subsection{Binomial observation model}
Our probabilistic model for read count data is based on the one used by PhyloSub\cite{phylosub}. Let $a_i$ and $b_i$ denote the number of reads matching the reference allele and the variant allele respectively at position $i$, and let $d_i = a_i+b_i$.  This represents the total number of reads at locus $i$. Let $\eta_k$ represent the population frequency of subpopulation $k$ (node $k$ in our tree). Let $\mu_i^r = 1-\epsilon$ denote the probability of sampling a reference allele from the reference population where $\epsilon$ is the error rate of the sequencer.  We set $\epsilon$ to $0.001$ for all our experiments.  Let $\mu_i^v$ denote the probability of sampling a reference allele from the variant population.  For the purposes of this paper we assume that all mutations are heterozygous and all loci have two copies, so we set $\mu_i^v$ to 0.5. Our model constrains the subpopulation frequencies $\eta_i$ such that $\eta_i \geq 0 \; \forall i$ and that $\sum_i \eta_i = 1$. We recover the node clonal frequencies using the recursive equation: $\phi_i = \eta_i + \sum_{j \in \mathcal{C}(i)} \phi_j$.
The observation model for read counts has the following likelihood:
\begin{equation}\label{eqn:pm}
\begin{aligned}
a_i \mid z_i =k, d_i,\phi_k,\mu^r_i,\mu^{v}_i  \sim \text{Binomial}(d_i,(1-\phi_k)\mu^r_i+\phi_k \mu_i^{v}) \ .
\end{aligned}
\end{equation}

\subsection{Inference}
Given this model and a set of $N$ observations $\{(a_i,d_i,\mu_i^r,\mu_i^v)\}_{i=1}^N$, the tree structure and the subpopulation frequencies $\{\eta_k\}_{k=1}^K$ are inferred using Markov Chain Monte Carlo (MCMC) sampling.  To sample the assignments of observations (SSMs) to nodes in the tree ($z_i$) we use Gibbs sampling where the probability of assigning a node is the prior probability (Equation \eq{eqn:tcrp}) multiplied by the likelihood of the data given that cluster identity (Equation \eq{eqn:pm}).  After completing a single pass through the observations in random order we then use Metropolis-Hasting sampling for 100 iterations to sample the $\eta_k$ variables.   We use an asymmetric Dirichlet distribution as the proposal distribution where the concentration parameter is set during burn-in to achieve an acceptance ratio between 0.05 and 0.75. 

For all experiments we sample for 2600 iterations and discard the first 100 samples as burn-in.

\subsection{Split-merge updates}
The Gibbs updates described previously only allow one cluster assignment to change at a time, which can result in slow mixing as described in the original treeCRP paper \cite{MeedsRZR08}.  Meeds \emph{et al.} \cite{MeedsRZR08} overcame this limitation by using split-merge updates in their implementation of the treeCRP.  However, their updates relied on the partial conjugacy of the cluster parameters as described in \cite{jain2004split}, which is not the case in our observation model.  In addition, the subclonal tree we are inferring has a natural ordering not present in the original treeCRP model.  This natural ordering is that the clonal frequency of a node in our tree must always be greater than or equal to the sum of the clonal frequencies of its children.  This natural ordering means that arbitrary split-merge moves are unlikely to be accepted.  We therefore propose two ``local'' split-merge updates that are more likely to be accepted: the \emph{parent-child split merge (PCSM)} and the \emph{leaf-sibling-split-merge (LSSM)}.

\begin{figure}[!t]
\centering
\begin{minipage}{.45\textwidth}
\includegraphics[width=1.\textwidth]{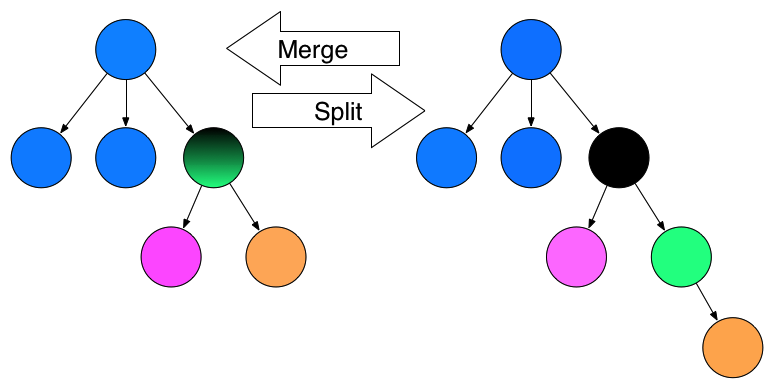}
\end{minipage}
\begin{minipage}{.45\textwidth}
\includegraphics[width=1.\textwidth]{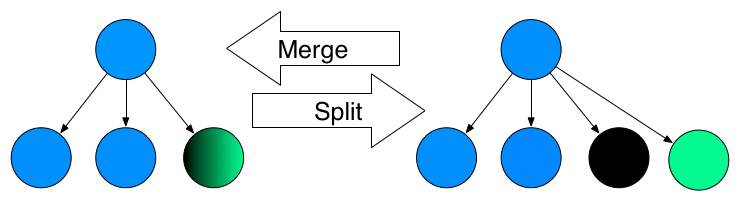}
\end{minipage}
\caption{Example of a Parent-Child Split-Merge move (\emph{left}) and a Leaf-Sibling Split-Merge move (\emph{right})}
\label{fig:sm}
\end{figure}

The PCSM selects a node in the tree at random and then with equal probability either splits or merges that node.
If merge is selected, then the node is merged with its parent (\emph{i.e.}, its SSMs are assigned to its parent) and all its children become its parent's children. If split is selected, a new node is added to the tree as the child of the selected node. The selected node's children are split with the new node, assigning a given child to the new node with
probability 0.5.  
The LSSM selects a leaf of the tree at random and, as the PCSM, selects with equal probability whether to add a new sibling node (i.e., split) or merge the selected node with a randomly selected sibling node.  Only leaf nodes are considered in this operation for implementation simplicity and because our subjective observation that leaf nodes exhibited slower mixing than the internal nodes. Whenever a new node is created through a split in either a PCSM or a LSSM update, the SSMs assigned to the split node are reassigned between the split and new node using restricted Gibbs updates.  The population frequency of a new node ($\eta_{new}$) is selected uniformly at random between 0 and the population frequency of its parent, while the parents population frequency is decreased by $\eta_{new}$ to maintain $\sum_i \eta_i = 1$. Figure \ref{fig:sm} shows an example of PCMS and LSSM updates. 

\section{Results}
In order to compare our approaches we constructed a series of simulated datasets and applied PhyloSub \cite{phylosub} (which uses the TSSB prior) and the treeCRP model with either Gibbs updates only (treeCRP-Gibbs), Gibbs updates and Parent Child Split-Merge moves (treeCRP-PCSM), Gibbs updates and Leaf-Sibling Split-Merge moves (treeCRP-LSSM) and all three types of updates (treeCRP-all). For treeCRP-all, we propose a PCSM update and then a LSSM update after each Gibbs update.  Our simulations looked at a range of total population counts (3, 4, 5, 6), read depths (20, 30, 50, 70, 100, 200, 300) and number of SSMs per population (5, 10, 25, 50, 100, 200, 500). In each case, the first population is a normal population with no associated SSMs, while each subsequent population is a descendant of all previous populations (i.e. a chain phylogeny).  For each simulated SSM $k$ in population $u$, reference allele reads ($a_k$) were drawn as:
\begin{align*}
a_k \sim \text{Binomial}(d_k, 1-\phi_u + 0.5 \phi_u) \ ; \quad 
d_k \sim \text{Poisson}(r) \ ,
\end{align*}
where $\phi_u$ is the clonal frequency of population $u$ and $r$ is the simulated read depth.
A table of the $\phi$ values used can be found as Table \ref{tbl:table1}.

\begin{table}[t]
\caption{Table of subclonal lineage proportions used}
\begin{center}
\begin{tabular}{|c|l|}
Number of populations & $\phi$ values used \\ \hline
3 & 0.44, 0.11\\
4 &	0.56, 0.25, 0.06\\
5 &	0.64, 0.36, 0.16, 0.04\\
6 &	0.71, 0.44, 0.25, 0.11, 0.03\\ 
\end{tabular} \label{tbl:table1}
\end{center}
\end{table}

\begin{figure}[!t]
\centering
\includegraphics[width=0.9\textwidth]{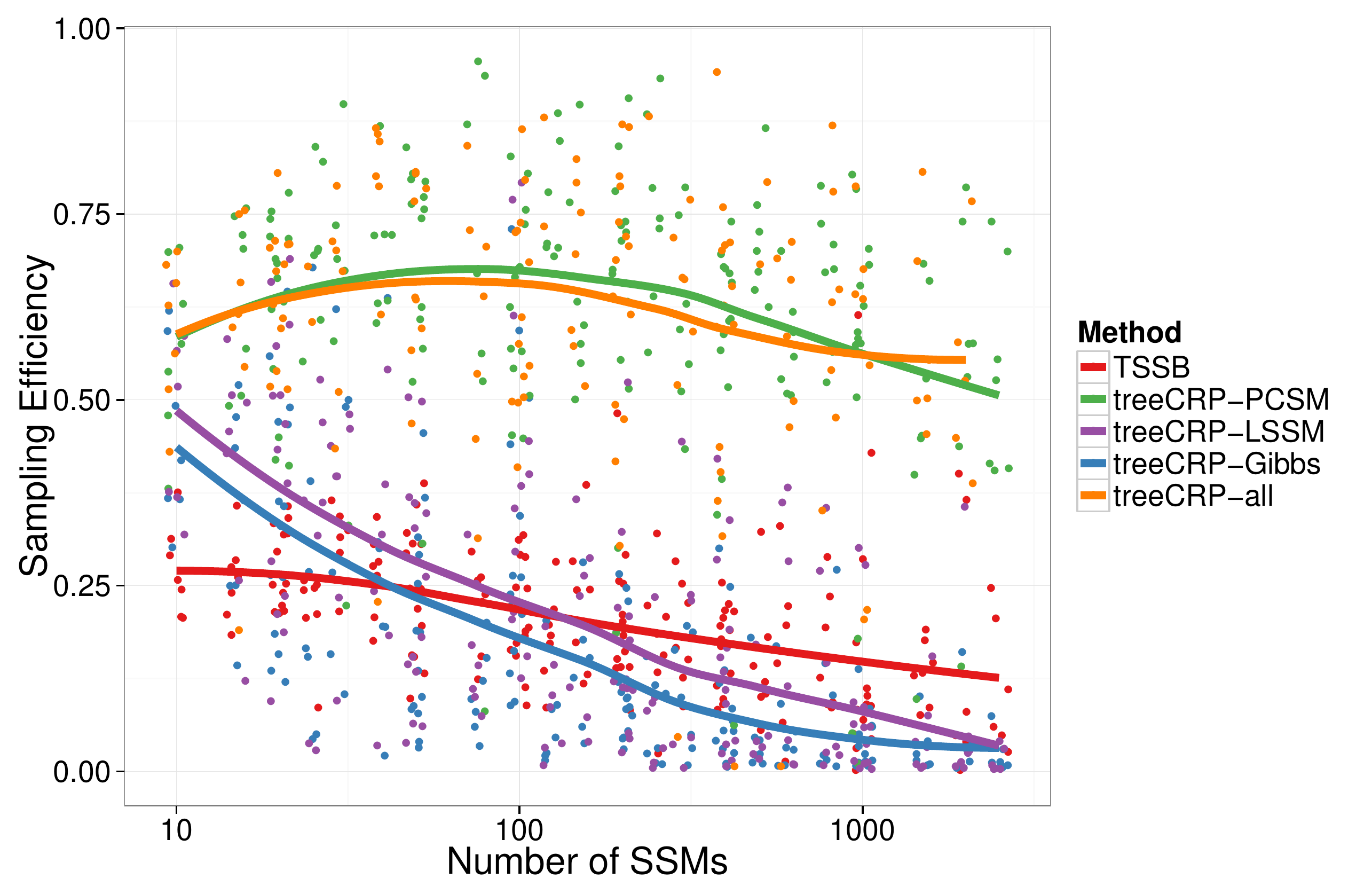}
\caption{The relationship between number of SSMs and sampling efficiency for the five algorithms.  Lines are are Loess curves. X-axis positions are jittered for clarity.}
\label{fig:ess}
\end{figure}

First, we compared how quickly the Markov chain mixes for the different tree priors and MCMC sampling strategies.  A chain that is fast-mixing requires fewer burn-in samples, is less likely to remain stuck in local modes and for a fixed desired accuracy requires fewer samples (or for the same number of samples delivers estimates with higher accuracy).  We measure mixing by first computing the effective sample size of the chain and then dividing by the number of actual samples taken. 
By dividing the effective sample size by the number of actual samples we get a measure of sampling efficiency, where a higher value indicates better mixing.  Effective sample size is calculated by the Coda package \cite{coda}, using the spectral density at frequency zero.
Figure \ref{fig:ess} shows the sampling efficiency for the five algorithms on our simulated datasets.  
The TSSB, the treeCRP-Gibbs and treeCRP-LSSM methods have consistently lower sampling efficiency than the treeCRP-PCSM and the treeCRP-all.
Furthermore, for treeCRP-PCSM and treeCRP-all the sampling efficiency is not strongly affected by the number of SSMs, whereas the efficiency of the other three methods decreases with increasing numbers of SSMs.

\begin{figure}[!t]
\centering
\includegraphics[width=0.9\textwidth]{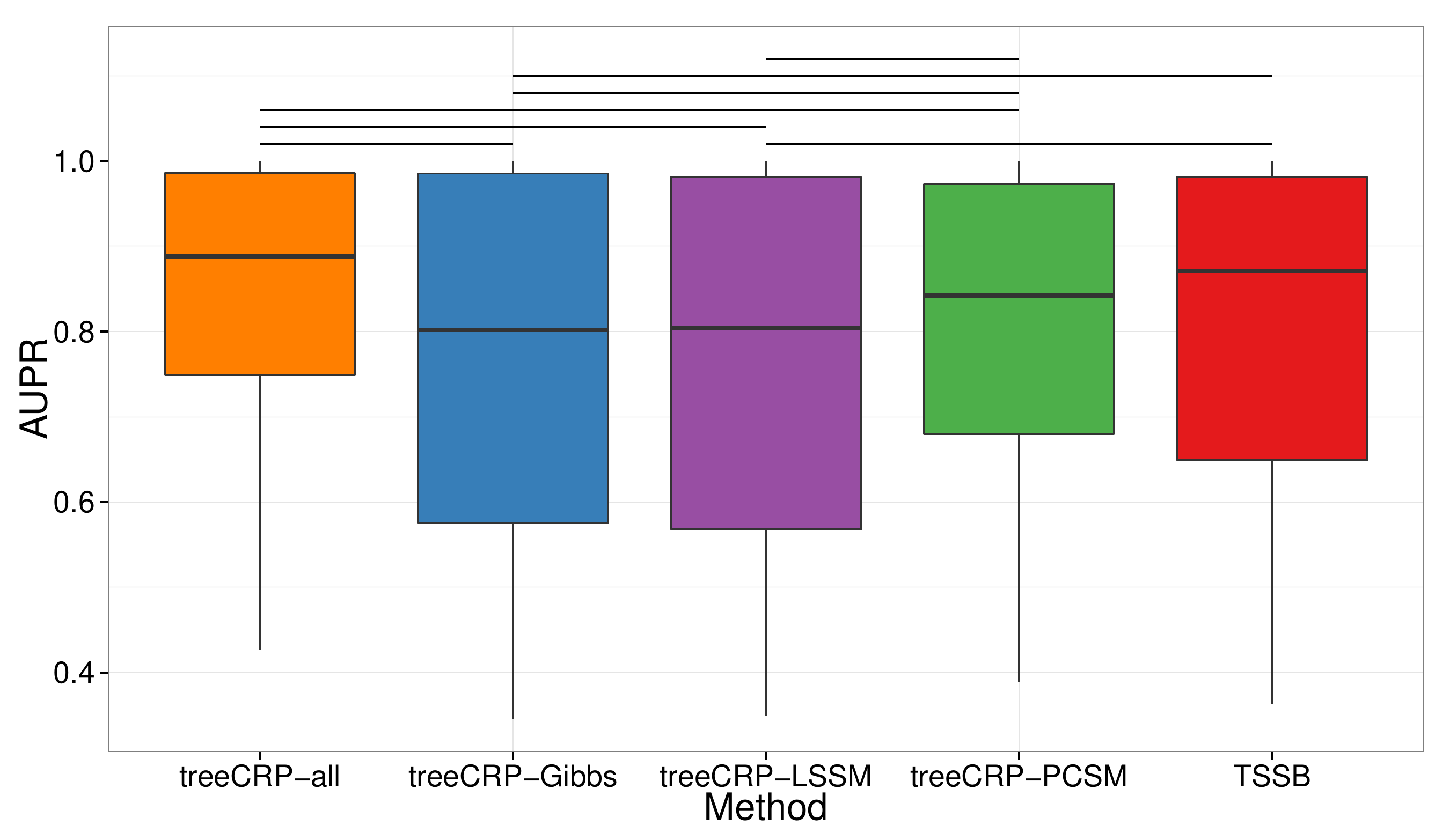}
\caption{Distribution of AUPR results for the five algorithms. Horizontal lines indicate statistically significant differences ($P < 0.005$, Wilcoxon paired signed rank test, Bonferroni correction)}
\label{fig:auprc}
\end{figure}

Next, we assessed the accuracy of the mapping from population to SSM.  To measure accuracy in a systematic way that accounts for class-imbalance, varying number of mutations and differing number of populations we used the Area Under the Precision-Recall Curve (AUPR) between the known true co-clustering matrix and the average co-clustering matrix over all samples. The co-clustering matrix $M$ is a binary matrix where $M_{ij} = 1$ if SSM $i$ and SSM $j$ are in the same subclonal lineage.
Figure \ref{fig:auprc} shows the distribution of AUPR values over all simulations. Although the absolute differences in AUPR are small, most of the pairwise differences are significant (7 out of 10, $P < 0.005$, Wilcoxon paired signed rank test, Bonferroni correction) and generally correspond to the differences in sampling efficiency in figure \ref{fig:ess} except that there are no significant differences between the AUPRs for TSSB and those of treeCRP-all and -PCSM. This suggests that the sampling efficiency differences have practical implications in reconstruction accuracy but neither prior is clearly superior.

\begin{figure}[!t]
\centering
\includegraphics[width=0.9\textwidth]{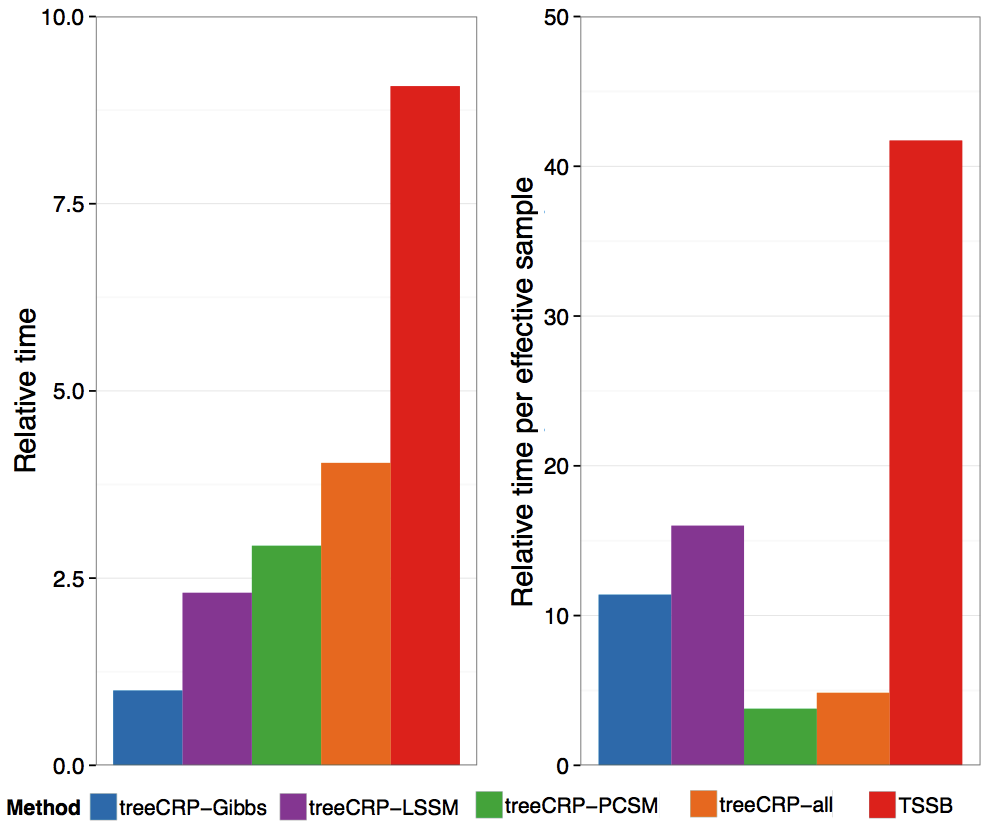}
\caption{Relative time per sample (i) and relative time per effective sample (ii) for the five methods}
\label{fig:runtime}
\end{figure}

Finally, we are interested in the relative time required to draw one sample.  This is because in most situations a sampling budget is a given amount of CPU time, so greater efficiency per sample could be counteracted by greater computational effort to compute that sample.  Because the cluster on which the experiments were run is composed of heterogeneous nodes it was meaningless to compare the runtimes of our experiments.  Instead, we ran all five algorithms on the same computer using the simulated dataset with 5 subpopulations, read depth of 200 and 500 SSMs per subpopulation.  Figure \ref{fig:runtime} (i) shows the average runtime per sampling iteration for the different algorithms, normalized to the runtime of the treeCRP-Gibbs algorithm while Figure \ref{fig:runtime} (ii) shows the runtime per effective sample.  We observe that the TSSB algorithm is the slowest followed by the -all, -PCSM, -LSSM and finally the Gibbs only variant of the treeCRP.  After adjusting for sampling efficiency, TSSB remains slower than the treeCRP methods but the treeCRP-PCSM and treeCRP-all are now about 5 times faster than the other treeCRP methods.  

\subsection*{Chronic Lymphocytic Leukemia}
\begin{figure}[!t]
\centering
\includegraphics[width=0.9\textwidth]{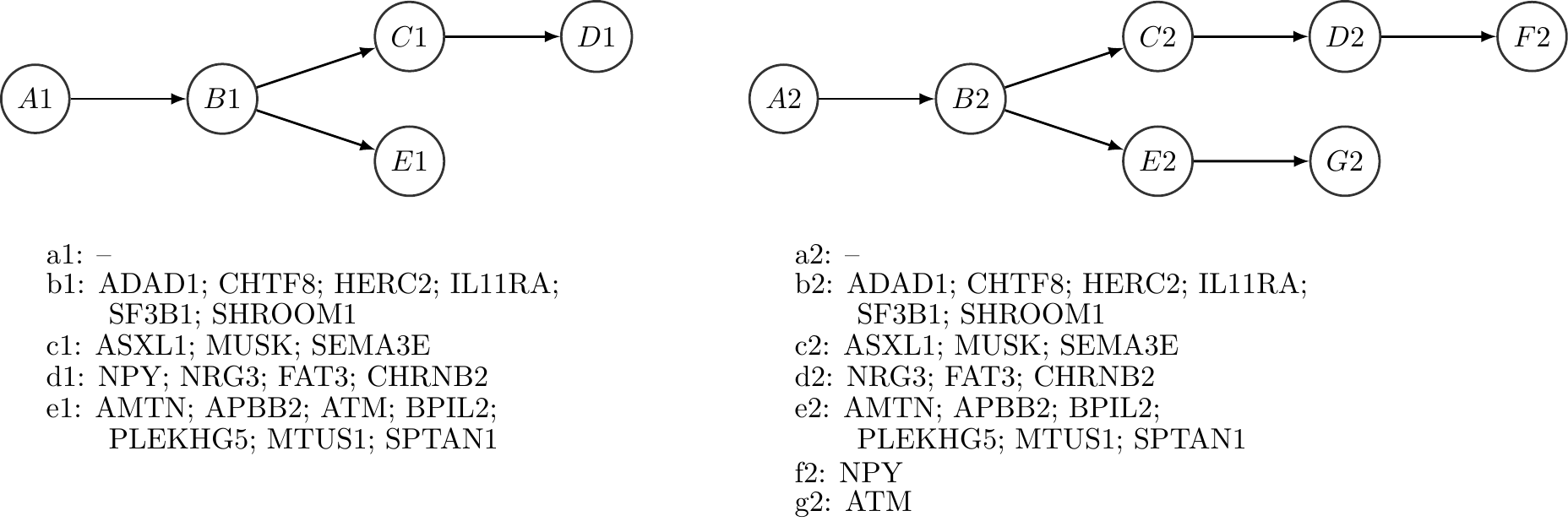}
\includegraphics[width=0.9\textwidth]{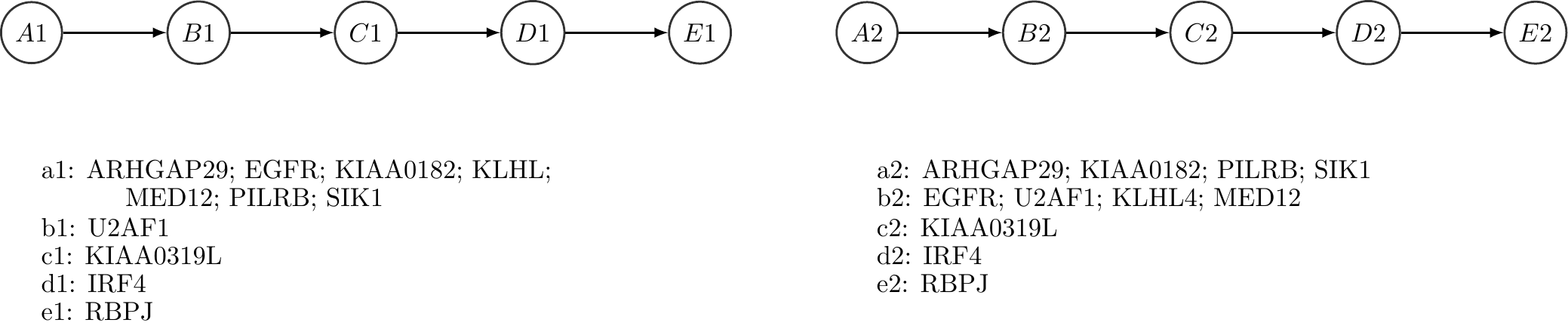}
\includegraphics[width=0.9\textwidth]{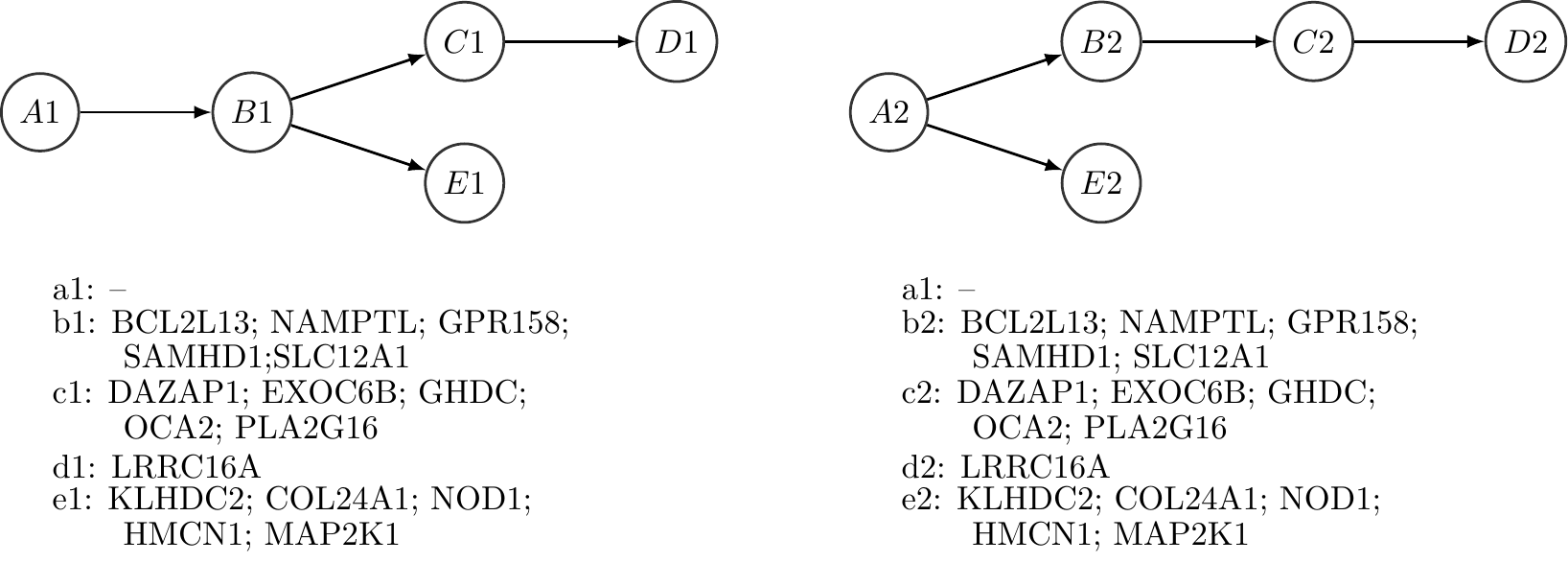}
\caption{Expert derived subclonal evolutionary trees (\emph{left}) and trees found by the treeCRP methods (\emph{right}) for patients CLL003 (top), CLL006 (middle) and CLL007 (bottom)}
\label{fig:cll}
\end{figure}

To demonstrate the ability of the treeCRP family of algorithms to perform subclonal reconstruction on a real dataset we applied the four methods to a a Chronic Lymphocytic Leukemia (CLL) dataset from Schuh \emph{et al.} \cite{Schuh12}.  The dataset consists of targeted sequencing data from three patients at five different time points; we reconstructed the tree using all five samples as input simultaneously.  We examined the maximum likelihood tree found during sampling.  All four treeCRP methods recovered the same tree structure and clustered the same SSMs together.  Figure \ref{fig:cll} shows the recovered tree structure along with the tree structure found in the original publication for the three patients CLL003, CLL006, and CLL077.  The trees we recovered are nearly identical to the expert derived trees, and those previously recovered by PhyloSub\cite{phylosub}, with a small number of differences. For the top two rows in fig. \ref{fig:cll}, the differences are minor changes in clonal frequency estimates that result in reassignment of some SSMs to direct parents or children. The change in the bottom row (CLL077) is more substantial, as the treeCRP methods are predicting that the E2 population is a direct descendant of normal rather than of the B2 population, in other words, there are two independent cancerous lineages. This change occurred from our previous reconstruction\cite{phylosub} because we no longer insist on a single cancer lineage in the new models. Although this reconstruction differs from the expert one, it is almost identical to one discovered by an independent, non-tree based method\cite{clomial}. 

\section{Conclusions}
In our experiments with simulated data the treeCRP prior delivered similar subclonal reconstruction accuracy to the TSSB while having reduced runtime per sample and per effective sample.
Among the treeCRP sampling strategies, treeCRP-all lead to the greatest subclonal reconstruction accuracy and second highest sampling efficiency among all five tested methods. 
When compared to the TSSB method, for the same amount of CPU time, the treeCRP-all method could generate 10 times the number of effective samples thus permitting a 10-fold decrease in run time. Furthermore, treeCRP-all's sampling efficiency was independent of SSM number whereas TSSB's decreased with larger numbers of SSM. So, treeCRP-all appears much more suited to subclonal reconstruction using whole genome sequencing data with tens of thousands of SSMs. 
However, surprisingly, despite the increase in effective number of samples, there was not a significant difference in reconstruction accuracy between treeCRP-all and TSSB. Furthermore, the TSSB reconstruction was a better match to the expert reconstruction on the CLL dataset.
As such, it remains an open question whether the decreased flexibility of the treeCRP prior (one hyperparameter versus two for TSSB) introduces a prior bias that interferes with subclonal reconstruction.

\section*{Acknowledgments}
This work was funded by a National Science and Engineering Research Council (NSERC) operating grant and an Early Researcher Award from the Ontario Research Fund to QM.
AGD is supported by a Natural Sciences and Engineering Research Council (NSERC) Vanier Canadian Graduate Scholarship.

\bibliographystyle{unsrt}
\bibliography{biblio}
\end{document}